\documentclass[twocolumn,notitlepage,showpacs,prl,superscriptaddress]{revtex4-1}

\usepackage{amsmath}
\usepackage{epsf}
\usepackage{graphicx}
\usepackage{dcolumn}
\usepackage{bm}

\newcommand{\req}[1]{Eq.~(\ref{#1})}
\newcommand{\reqs}[1]{Eqs.~(\ref{#1})}
\newcommand{\rref}[1]{(\ref{#1})}

\newcommand{\pt}{\partial_t}
\newcommand{\beq}{\begin{equation}}
\newcommand{\eeq}{\end{equation}}
\newcommand{\be}{\begin{equation}}
\newcommand{\ee}{\end{equation}}
\newcommand{\beqa}{\begin{eqnarray}}
\newcommand{\eeqa}{\end{eqnarray}}
\newcommand{\bea}{\begin{align}}
\newcommand{\eea}{\end{align}}
\usepackage{amssymb}
\usepackage{array}
\usepackage{amsmath}
\usepackage{graphicx}\usepackage{graphics}
\usepackage{dcolumn}
\usepackage{bm}\usepackage{varioref}

\newcommand{\he}{\hat{\epsilon}}

\begin{document}
\def\boldsymbol#1{\mbox{\boldmath$#1$}}

\title{Non-analytic Vortex Core and a Nonlinear Vortex Flow in Bosonic Superfluids}

\author{Oded Agam}
\affiliation{The Racah Institute of Physics, The Hebrew University of Jerusalem, 91904, Israel}
\author{Igor L. Aleiner}
\affiliation{Physics Department, Columbia University, New York, NY 10027, USA}

\date{\today}
\begin{abstract}
We analyze the disorder limited motion of  quantum vortices
in a two-dimensional bosonic superfluid with a large healing length.
It is shown that the
excitations of low-energy degrees of freedom
associated with the non-analytic reconstruction of the vortex core
[Ann. Phys. {\bf 346}, 195 (2014)] determine
strong non-linear effects in the vortex transport at velocities much
smaller than Landau's critical velocity.  Experiments are suggested
to verify our predictions.
\end{abstract}

\maketitle
{\em Introduction}--
Entropy production and dissipation in superfluids and
superconductors are associated with the dynamics of vortices. The cores of these vortices play a central
role in these processes. For instance, the Bardeen-Stephen friction force acting on a vortex in type
II superconductors comes essentially from the current flowing through
its normal core \cite{BardeenStephen65}. Beyond the linear response
regime, Larkin and Ovchinnikov showed that the quasiparticle
distribution within the vortex core deviates from the equilibrium
distribution due to the long inelastic relaxation time.
The heating  in the core of the vortex leads to a nonlinear behavior of the
friction force acting on the vortex even when the value of the applied current
is much smaller than the critical one \cite{LarkinOvchinnnikov75,LarkinOvchinnikovReview,VortexReview}.

There is no analogous scenario for vortex flow in bosonic superfluids. This is
 because, unlike superconductors (and $^3$He fermionic superfluid),
quantum vortices in such a superfluid were thought to posses an
essentially featureless cores, see e.g. Ref.~\cite{Review83}.

\begin{subequations} \label{eq:1}
However, we showed recently \cite{KleinAleinerAgam2014} that vortices
in two-dimensional bosonic superfluids experience non-analytic reconstruction of their cores when moving
with respect to the flow. The theory \cite{KleinAleinerAgam2014}
developed for the   limit, $n\xi^2 \gg 1$, where $n$ is the bosonic density
and $\xi$ is the healing length (i.e. the size of the vortex core), predicts the following:
i) The low energy degrees of freedom are not exhausted by the position of the vortex itself but must include
 the precession of the vortex around its guiding center; ii) This precession can be characterized by the
kinetic momenta $\hat{\vec{p}}=(\hat{p}_{x},\hat{p}_y)$ such that
\be
\left[\hat{p}_{x},\hat{p}_{y}\right]=i2\pi\sigma n\hbar^2,
\label{eq:1a}
\ee
where $n$ is the bosonic density far from the vortex core, and $\sigma=\pm 1$ is the vorticity,
 \req{eq:1a} corresponds to the Lorentz (Magnus) force acting on a moving quantum vortex with respect to the  superfluid;
iii) The momentum dependence of the kinetic energy is non-analytic
\be
\hat{H}_{k}(\hat{\vec{p}})=\frac{\hat{\vec{p}}^2}{2M_v(\hat{\vec{p}}^2)};
\  \frac{M_v(p^2)}{m n\xi^2}\equiv
\frac{\pi}{\alpha^2} \ln\left(\frac{\hbar^2 n^2\xi^2}{p^2} \right),
\label{eq:1b}
\ee
where $m$ is the boson mass, and $\alpha=0.802\dots$;
iv) Semiclassical quantization $\hat{H}_{k}(\hat{\vec{p}})$ gives discreet energy levels
\be
\epsilon_{l+1}-\epsilon_{l}={\hbar\omega_c^l};
\quad
\omega_c^l=\frac{2\alpha^2\hbar}{m\xi^2}\left[ \ln\left(\frac{\hbar\omega_c^l n\xi^2}{\epsilon_{l}} \right)\right]^{-1},
\label{eq:1c}
\ee
where $l\geq 0$ is an integer; v) Excited states ($l\geq 1$) decay due to the phonon emission
but the relaxation time, $\tau_{in}$, is large and discrete levels are distinguishable
 \be
 \frac{1}{\tau_{in}^l} =\frac{2\pi\alpha^8\hbar}{m \xi^2} \left[ \ln
\left(\frac{\hbar\omega_c^l n\xi^2}{\epsilon_{l}} \right)\right]^{-4}.
\label{eq:1d}
 \ee
The spectrum \rref{eq:1b} with vortex mass $M_v\simeq mn\xi^2$ was discussed extensively in the literature
concluding that the excited states of the vortex are not relevant for the low-energy dynamics.
The large logarithmic factor in the mass, see Eq.~(\ref{eq:1b}), makes the dynamics much slower than it was previously thought \cite{KleinAleinerAgam2014}.
\end{subequations}

\begin{figure}[h]
\includegraphics[width=0.6\columnwidth]{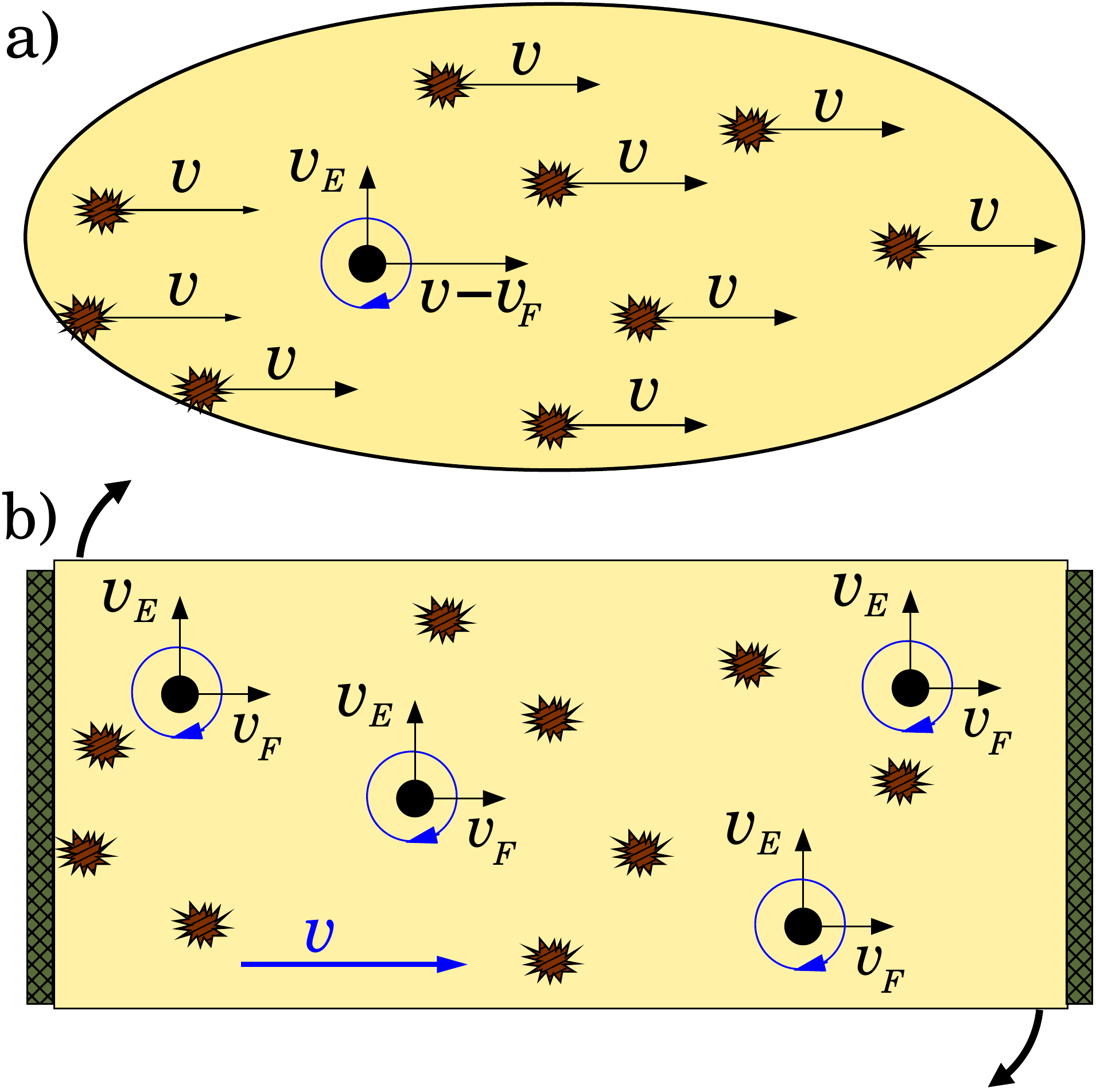}

 \caption{Setups for a) cold atom (BEC)
 based and b) helium films superfluids.
 In BECs, vortex (black dot) can be introduced
 by phase imprinting \cite{phaseimprinting}, while motion is induced
by sweeping (with velocity $v$) a disordered potential generated by a speckle pattern (brown impurities) \cite{movingdisorder}.
 For  helium films vortices can be introduced by rotating the sample \cite{Review83}
(so that the vortex lattice is not yet formed),
and the superfluid motion is obtained by evaporative heater located
at one end of the system and a reservoir of superfluid helium at the other end .
}
\label{fig0}
\end{figure}

Whereas a direct observation of the  core
dynamics is difficult, its manifestation via the dissipative motion of vortices,  like those in superconductors
\cite{LarkinOvchinnnikov75,LarkinOvchinnikovReview},
is experimentally accessible, as we demonstrate in this Letter.

In a stationary clean superfluid the vortices move with the flow and
the internal degrees of freedom are not excited.
The experiments to pinpoint the excitations in disordered systems are sketched in Fig.~\ref{fig0}.  For a BEC
[Fig.~\ref{fig0}a)] the motion of the vortices can be observed {\em in
  situ} whereas for the helium film realization [Fig.~\ref{fig0}b)] it
can be deduced from measurements of the chemical potential gradient along the superfluid flow by means of differential pressure transducer \cite{C3}. The superfluid flow is obtained by evaporative heater located at one end of the system and a reservoir of superfluid helium at the other end \cite{C2}. Vortices with the same vorticity are induced by mounting the sample in a rotating cryostat, see e.g. Ref.~\cite{C1}.

\begin{figure}[h]
\includegraphics[width=0.9\columnwidth]{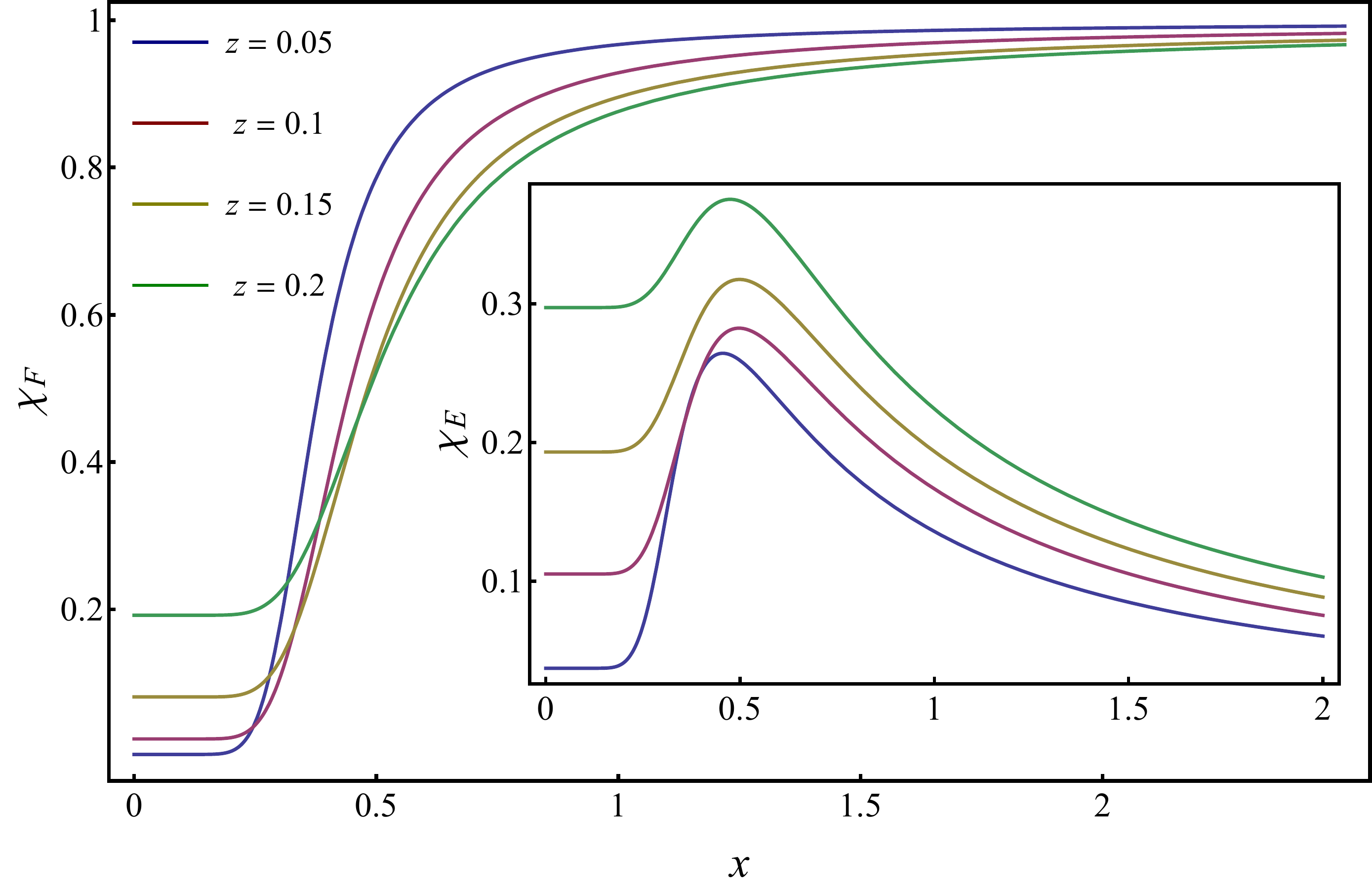}
\caption{Non-linear susceptabilities as the function of the velocity $v$ expressed in terms of the ``threshold'' velocity $v^*$ determined
by the disorder and the phonon temperature $x\equiv v/v^*$. Different curves correspond to different phonon temperature $T$ expressed
in units of the disorder energy $\epsilon_d$, see \req{eq:4a}, $z=T/\epsilon_d$. Naturally, the threshold behavior becomes more pronounced with the lowering of the temperature.}
\label{fig1}
\end{figure}

The main quantities are
the non-linear susceptabilities:
\be
v_F=\chi_F(v)v; \quad v_E=\sigma \chi_E(v)v,
\label{chi}
\ee
where the vortex, $v_{E,F}$, and superfluid velocities, $v$, are introduced in Fig.~\ref{fig0}
($v$'s are to be understood
as averages over the disorder realizations).
Our predictions for the susceptabilities
are summarized in Fig.~\ref{fig1} and  the analytic
expressions are given at the end of this paper.

In BEC systems, the motion of the vortex can be imaged
 and $\chi_{F,E}$ are directly measurable.  For  helium films the chemical potential gradient, $\vec{\nabla} \mu$, satisfies the relation
\be
\vec{\nabla} \mu=-2\pi\hbar N_V \chi_E(v) v,
\label{alpha}
\ee
 where $N_V$ is the density of the vortices per unit area. Thus $\chi_E(v)$ can be extracted from the ratio of the chemical potential gradient to the superfluid velocity. Moreover,  the N-shape of $\chi_E(v)v$ leads to the situation where the same vortex current can be realized for three possible value of the supercurrent $v$, thus leading to the filament instability of the supercurrent itself in large samples.

{\em Qualitative discussion} --
To discuss the motion of the vortex it is necessary to supplement the kinetic energy
\rref{eq:1b} with the fields coming from the motion of the superfluid surrounding the vortex.
The corresponding effective theory  \cite{KleinAleinerAgam2014} is conveniently written
within Popov's formalism \cite{Popov83}
mapping the problem of two-dimensional superfluid to two-dimensional nonlinear electrodynamics.
In this mapping, vortices become charged
particles with charge $\sigma 2\pi\hbar$, the electric field $E$ is related to the superfluid
current $\vec{j}$ as $\vec{E}=-\hat{\epsilon}\vec{j}$ (where $\hat{\epsilon}$ is the
antisymmetric tensor of the second rank acting on the spatial
coordinates), and the magnetic field, $B$, is the boson density, $n$.
In Popov's variables, the effective Hamiltonian of a vortex (which includes the reconstruction of the core) is
\be
\hat{\cal H}=
H_k\left( \vec{p}\right) + 2\pi\sigma\hbar\, \varphi + \frac{\vec{p}\,\hat{\epsilon}\,\vec{ E}}{B} +  \frac{\pi \hbar^2 B(\vec{r})}{m},
\label{eq:2}
\ee
where $\vec{p}= \vec{\cal P} - 2\pi\sigma \hbar \vec{ A}(\vec{r},t)$ is the kinetic momentum of the vortex \rref{eq:1a},
while $\vec{\cal P}$ is the canonical one, $\varphi(\vec{r},t)$ and $\vec{A}(\vec{r},t)$ are the scalar and the vector potentials,
 respectively. The physical  fields are $\vec{E}= -\vec{\nabla}\varphi- \partial_t \vec{A}$ and $B= \vec{\nabla} \times \vec{A}$.
The gauge invariance of \req{eq:2} is nothing but vorticity conservation.

The third and the fourth terms in \req{eq:2} are not present in usual electrodynamics.
The third term expresses the fact the vortex executes its motion around the guiding center moving
together with the superfluid and manifests Galilean invariance. The last term is the energy of the core depending on the superfluid density outside the core,
$B(\vec{r})$. We will see shortly that this term is important for the scattering of the vortex by disorder.

It is instructive to write the equation of motion from the Hamiltonian \rref{eq:2}.
Suppressing  the non-important effects of the spatial inhomogeneity of $E/B$, we have
\begin{subequations}
\label{eq:3}
\beqa
\dot{\vec{p}}&=&2\pi\sigma\hbar \left( \vec{E} + B \hat{\epsilon} \dot{\vec{r}} \right)-{\pi\hbar^2 \vec{\nabla} B(\vec{r})}/{m}
\label{pdot} \\
\dot{\vec{r}}&=&\partial_{\vec{p}} \hat{H}_{k}(p) + \hat{\epsilon}{ \vec{E}}/{B}. \label{rdot}
\eeqa
and combining \reqs{pdot}--\rref{rdot}, we find
\be
\dot{\vec{p}}=\pm \omega_c(p^2)\he\vec{p} -{\pi\hbar^2 \vec{\nabla} B(\vec{r})}/{m},
\ee
\end{subequations}
{\em i.e.}  the electric field cannot excite the external
degrees of freedom of the vortex.
Such excitations can be induced only
by scattering on inhomogeneities of the bosonic density that we will describe now.

The analogy of the problem with the motion of electron in magnetic field enables us to use the
formalism developed for the nonlinear magnetotransport in a two dimensional electron gas \cite{VavilovAleiner}.
Consider the small Gaussian variations in the boson density $n=n_0+\delta n$,
\be
\langle \delta n(\vec{r}_1)\delta n(\vec{r}_2)\rangle_q=\gamma n_0G(qr_c),
\label{eq:4}
\ee
where $\langle \dots\rangle_q $ denotes the disorder averaging and
Fourier transform over $\vec{r}_1-\vec{r}_2$, the parameter $\gamma \ll 1$ describes the disorder strength, $r_c$
is the correlation radius  \cite{potential}, and $G(x)$ is the dimensionless function whose precise form is not important for us provided that it drops fast enough at $x \gg 1$.
From \req{eq:1a} it follows that $p^2 \geq \hbar^2 n_0 \gg \hbar^2/r_c^2$ which means that the vortex
experiences small angle scattering by the disorder. The relaxation times can be
estimated by treating the last term in \req{eq:2} within the
Fermi golden rule and neglecting the curving of the trajectories between scattering events.
It gives the following relaxation time for the momentum direction:
\begin{subequations}
\be
{1}/{\tau_{tr}(\epsilon)}=\omega_c\left({\epsilon_d}/{\epsilon}\right)^{3/2},
\label{eq:4a}
\ee
where $\epsilon_d$ is the characteristic energy scale \cite{footnote1} associated with the disorder.
If the kinetic energy  of the vortex $\epsilon$ is larger than $\epsilon_d$ the vortex
precesses many times before changing its position, otherwise
the vortex scatters into the new position before it manages to complete the circle.

There is another time scale describing the scattering at all angles (and not only those
which change the direction of the momentum significantly) \cite{footnote1}:
\be
{1}/{\tau_q(\epsilon)}=\omega_c\left({\epsilon_q}/{\epsilon}\right)^{1/2}
,\quad
\epsilon_q \gg \epsilon_d.
\ee
If $\omega_c\tau_q\gtrsim 1$, one can neglect the interference associated with the coming back to the same scattering center (Shubnikov-de-Haas effect). We will assume $\epsilon < \epsilon_q$, whereas the relation between $\epsilon_d$ and $\epsilon$ may be arbitrary.
\end{subequations}

Consider now a vortex in the coordinate frame moving with velocity $\vec{v} = \he \vec{E}/B$ [moving disorder
in this frame does not change $H_k$ because of the last term in \req{eq:2}].
To be at rest in the laboratory frame (which would be consistent with disorder pinning), a vortex should have the directed velocity  $-\vec{v}$.
If there were no disorder,  $\tau_{tr}\to \infty$, this directed velocity would precess and average to zero.
The presence of disorder allows for rotation by small angle $\omega_c\tau_{tr} \ll 1$. As a result the velocity acquires a component along $E$,
$v_E\simeq \sigma \omega_c\tau_{tr}\vec{E}/B$. The motion along the electric field leads to Joule heating,  and
the power produced by  vortices with typical energy $\epsilon <\epsilon_d$ [see \req{eq:4a}] is
\[
 P(\epsilon)/(2\pi\hbar)\simeq\sigma v_EE\simeq \omega_c\tau_{tr}{\vec{E}^2}/{B}
\simeq   {\vec{E}^2}/{B}\left({\epsilon}/{\epsilon_d}\right)^{3/2}\!\!\!\!\!.
\]
In the opposite limit $\omega_c\tau_{tr}\gg 1$, the circular motion averages out
the dissipative current, and only rare scattering events contribute to the dissipation power. Therefore the dissipative current should be proportional to $1/\tau_{tr}$, and on dimensionality grounds it leads  the
replacement, $\omega_c\tau_{tr} \to 1/(\omega_c\tau_{tr})$, i.e. for ${\epsilon} \geq {\epsilon}_d$,
\[
 P(\epsilon)/(2\pi\hbar) \simeq { \omega_c\tau_{tr}}{\vec{E}^2}/{B}
\simeq {\vec{E}^2}/{B}\left({\epsilon_d}/{\epsilon}\right)^{3/2},
\]

\begin{figure}[h]
\includegraphics[width=0.7\columnwidth]{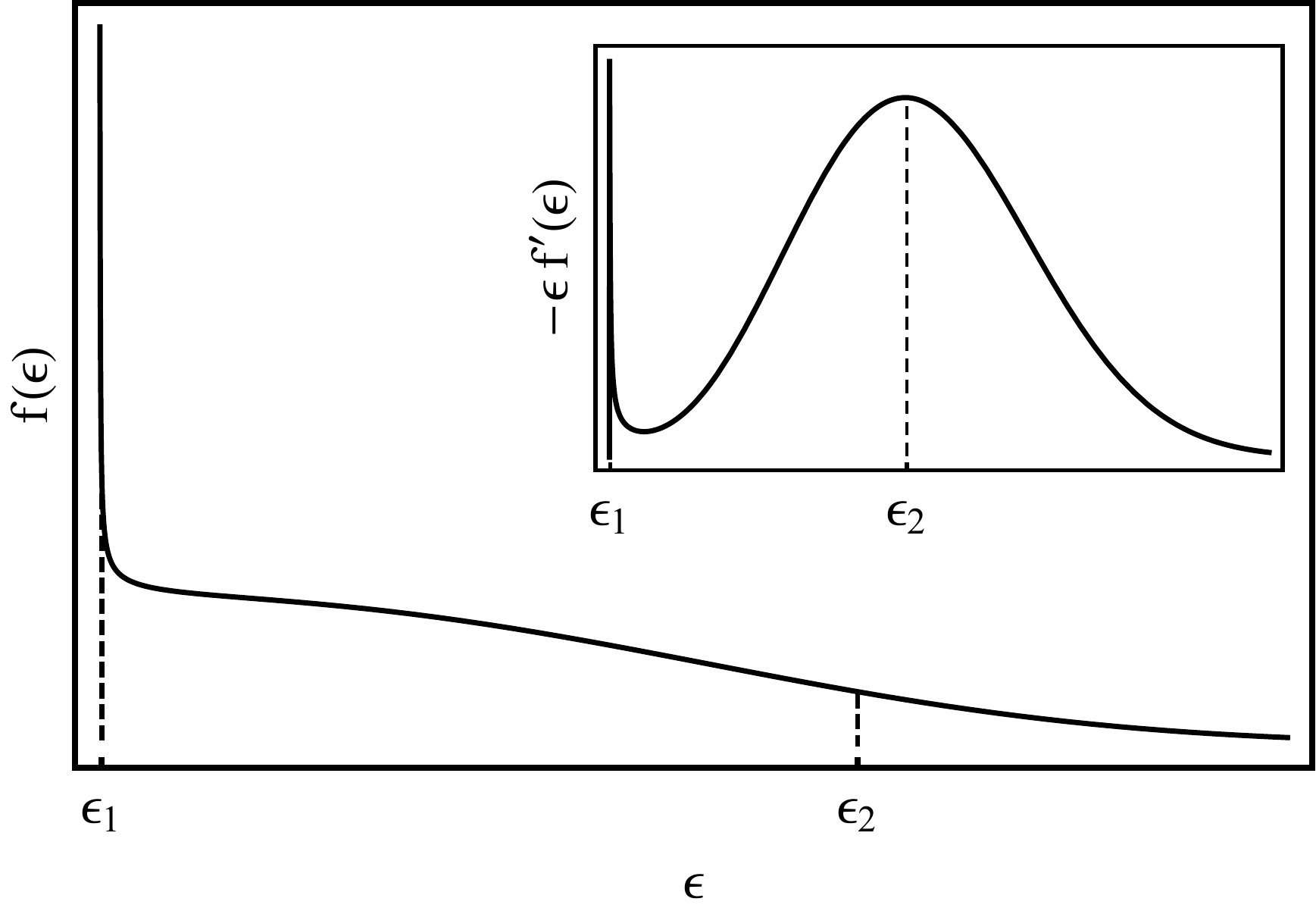}
\caption{The shape of the non-equilibrium distribution function $f(\epsilon)$ calculated from \req{distribution}.  The steep drop at low energies is the thermal distribution and the high energy
tail originates for the energy resolved ``heating'' \rref{Delta}. For $\epsilon_1 < \epsilon < \epsilon_2$, the relation $\Delta(\epsilon) >\epsilon $ holds.
 }
\label{Fig2}
\end{figure}

Thus the generation of energy is a peaked function of $\epsilon$, and non-equilibrium effects are
associated with its particular form. If there were no inelastic processes
the distribution function of the vortices in the energy space $f(\epsilon)$ would never be stationary.
Phonon emissions \rref{eq:1c} remove the energy from the vortex core, and the extra energy accumulated by the vortex with reference to the starting energy $\epsilon$ can be estimated as
\be
\Delta(\epsilon) \simeq  \left({2\pi\hbar \tau_{in} E^2}/{B}\right)
\left[
\left(\epsilon_d\,\epsilon\right)^{3/2}/(\epsilon_d^3+\epsilon^3)\right].
\label{Delta}
\ee
Therefore, for large enough $E$, there exists a region where $\Delta(\epsilon) \gtrsim \epsilon$.
The distribution function in such a region is almost constant, see
Fig.~\ref{Fig2}. Non-equilibrium currents, however, are determined by the energy derivative of  the distribution function $(-\partial f/\partial \epsilon)$,
shown in inset of Fig.~\ref{Fig2}.
Thus, the currents are not determined by the whole distribution function but only by
regions at small and  large energies where the dissipative currents are suppressed.
This explains the non-linearity of dissipation as function of $E/B$, and the drop in the dissipative current
at large $E/B$. The quantitative qualitative requires full kinetic description of the problem outlined below.

{\em The kinetic equation} has the standard form in the energy-angle variables
\cite{Sorry}. Suppressing the spatial dependence of the distribution function,
 $f(\epsilon,\phi)$, we obtain
\begin{subequations}
\be
\pt f
+\sigma\omega_c(\epsilon) {\partial_\phi f}
= {\mathrm St}_{el} [f]+{\mathrm St}_{in}[f].
\label{kineqn}
\ee
The collision integrals in the right-hand-side of \req{kineqn} describe probabilistic processes.
The disorder generates small angle scattering
(angular diffusion):
\be
{\mathrm St}_{el} [f]
\! = \! \left[\frac{ \partial}{\partial \phi }- \frac{\partial}{\partial \epsilon }\frac{\vec{E}\cdot \vec{p}}{B}\right]
\frac{1}{\tau_{tr}}
\left[\frac{ \partial}{\partial \phi }-\frac{\vec{E}\cdot \vec{p} }{B} \frac{\partial}{\partial \epsilon }\right]
 f,
\label{St-el}
\ee
where the elastic relaxation time $\tau_{tr}(\epsilon)$ is  given by \req{eq:4a}.
We defined the angle $\phi$ so that $\vec{E}\cdot \vec{p}=Ep(\epsilon)\cos\phi$, and  $p(\epsilon)=\sqrt{2\epsilon M_v(\epsilon)}$, see \req{eq:1b}. The extra term in addition to the angular
derivative is the Galilean correction to the vortex energy in the moving superfluid [third term in \req{eq:2}].
The energy transfer in the phonon emission  is small and can be described by Focker-Planck terms.
 Neglecting the effects of the field $E$ and the disorder  on the inelastic collision, we obtain
\be
{\mathrm St}_{in}[f]
= \frac{ \partial}{\partial \epsilon} \left[
\frac{\epsilon}{\tau_{in}(\epsilon)}
\left( 1  + T   \frac{ \partial}{\partial \epsilon} \right)\right] f,
\label{St-in}
\ee
where $T$ is the phonon temperature, and the inelastic rate is given by \req{eq:1c}.
\label{kinetics}
\end{subequations}

The vortex velocities
(in the convention of Fig.~\ref{fig0})
are
\be
v_E\!=\!\int\! f\frac{\partial \epsilon}{\partial p}\cos\phi\, d\phi d\epsilon;
\,  v_F\! =\frac{E}{B}+
\!\int\! f\frac{\partial \epsilon}{\partial p}\sin\phi\, d\phi d\epsilon,
\label{currents}
\ee
with the normalization $\int d\phi d\epsilon f=1$.

{\em Solution of the kinetic equation} proceeds in a standard way
{\em e.g.} considering the heating effects in metals.
Let  $f(\epsilon;\phi)=f_0(\epsilon)+f_1(\epsilon;\phi)$,
and $\int d\phi f_1(\epsilon;\phi)=0$.
The angular dependent part of the distribution
function $f_1$ is massive and can be found to first order
in the perturbation in $E$. For the same reason, inelastic collision
effects on $f_1$ can be also neglected, and we find
\be
f_1
=\frac{E p(\epsilon)}{B}\
\frac{\sigma\omega_c(\epsilon)\tau_{tr}(\epsilon)\cos\phi - \sin\phi
}{1+\omega_c(\epsilon)^2\tau_{tr}(\epsilon)^2}
\left(-\frac{\partial f_0}{\partial\epsilon}\right).
\label{f1}
\ee
Substituting $f(\epsilon;\phi)=f_0(\epsilon)+f_1(\epsilon;\phi)$
back into \req{kineqn}, using \req{f1} and integrating
the result over the angle $\phi$ we obtain the spectral
diffusion equation $\partial_tf_0+\partial_\epsilon j_\epsilon=0$,
where the spectral flow current is given by
\begin{subequations}
\be
j_\epsilon=-{\epsilon f_0(\epsilon)}/{\tau_{in}(\epsilon)}
- D_\epsilon\partial_\epsilon f_0(\epsilon).
\ee
The spectral diffusion is caused both by the inelastic processes and by
Joule heating due to the electric field:
\be
 D_\epsilon  =
\frac{\epsilon\, T }{\tau_{in}(\epsilon)}
+\frac{1}{2} \left(\frac{E p(\epsilon)}{B}
\right)^2\frac{\omega_c(\epsilon)^2\tau_{tr}(\epsilon)}
{1+\omega_c(\epsilon)^2\tau_{tr}(\epsilon)^2}.
\ee
\label{spectralflow}
\end{subequations}
In the stationary state the spectral flow is absent $j_\epsilon=0$ and we obtain from \reqs{spectralflow}
\be
\frac{\partial f_0}{\partial\epsilon }+\frac{f_0}{T_{eff}(\epsilon)}=0;\
T_{eff}=T+\frac{2\pi\hbar \tau_{in} E^2}{B}\frac{\left(\epsilon_d\,\epsilon\right)^{3/2}}{\epsilon_d^3+\epsilon^3}
\label{Teff},
\ee
where we used $p^2=2M_v\epsilon$, $\omega_c=2\pi\hbar B/M_v$, and the explicit energy depedence of the
transport relaxation rate \rref{eq:4a}. The meaning of the last term in the expression for the effective temperature
$T_{eff}$ has been already discussed in derivation of \req{Delta}.

Equations \rref{currents} and \rref{f1} give
\be
\begin{bmatrix}
v_E\\
v_F
\end{bmatrix}
= \frac{E}{B}
\int_0^\infty
\begin{bmatrix}\pm
\left(\epsilon_d\ \epsilon\right)^{3/2}\\
\epsilon^3
\end{bmatrix}
\left(-\frac{\partial f_0}{\partial \epsilon}\right)\frac{\epsilon d\epsilon}{\epsilon_d^3 + \epsilon^3}.
\label{currents-formula}
\ee
The normalized solution of \req{Teff} is
\be
f_0(\epsilon)=\frac{F(\epsilon)}{\int_0^\infty d\epsilon_1F(\epsilon_1)};
\ F=\exp\left[-\int_0^\epsilon \frac{d\epsilon_1}{T_{eff}(\epsilon_1)}\right].
\label{distribution}
\ee
Substituting \req{distribution} into \req{currents-formula}, restoring $v=E/B$, and matching overlapping
asymptotes for the integrals we obtain for the susceptabilites $\chi_{\nu=E,F}$ of \req{chi}:
\begin{subequations}
\label{final}

\be
 \chi_{\nu}
 =\frac{\left(\frac{2T}{\epsilon_d}\right)^{\delta_\nu^T}d_{\nu}^T+\left(\frac{v}{v^*}\sqrt[4]{\frac{\epsilon_d}{T}}\right)^{\delta_\nu^v}
e^{-\left({v^*}/{v}\right)^{4/3}} d_{\nu}^v
}
  {\left(\frac{T}{\epsilon_d}\right) +
 \left(\frac{v}{v^*}\sqrt[4]{\frac{\epsilon_d}{T}}\right)^{\delta_F^v}e^{-\left({v^*}/{v}\right)^{4/3}}  d^v_F},
\label{finalchi}
 \ee
 where the exponents are
 $\delta_E^v=-2/5;\ \delta_F^v=4/5;\ \delta_E^T=5/2;\ \delta_F^T=4 $,
and the numerical prefactors are all of the order of unity:
$d^T_E=0.58\dots,\ d^T_F=3/2,\ d^v_E= 0.583\dots,
\  d^v_F=3.53\dots
$.
The nonlinearity occurs at ``threshold'' velocity
\be
\frac{v_*}{c}=\left[\frac{\beta}{ n_0\xi^2}\frac{m}{M_v}\frac{\hbar\omega_c}{T}\right]^{1/4}
\!\left[\frac{m\xi^2\epsilon_d}{\hbar^2}\right]^{3/4}\!
\left[\frac{1}{\omega_c\tau_{in}}\right]^{1/2}\!\!\! ,
\label{v*}
\ee
where $\beta=32\pi^2/(81\sqrt{3})=2.25\dots$, and
$c=\hbar/(m\xi)$ is the speed of sound.
With the logartithmic accuracy, one uses $\epsilon_l\to T \gtrsim \hbar\omega_c$
in expressions \rref{eq:1c}-\rref{eq:1d}. At smaller temperatures, one should replace
$T \to \hbar\omega_c(\epsilon_l=\hbar\omega_c)$. Each fraction in \req{v*} is
small so that the non-linearity occurs at a superfluid velocity much
smaller than Landau's critical value.
This value of $v_*$ can be understood from the condition $T_{eff}(\epsilon)=\epsilon=2T$, whose meaning
is obvious from the qualitative discussion and Fig.~\ref{Fig2}.
\end{subequations}

{\em In conclusion}, we constructed the theory for the motion of quantum vortices in disordered
two-dimensional bosonic superfluids. The excitations of low energy degrees of freedom,
associated with  core reconstruction \cite{KleinAleinerAgam2014},
lead to non-linear transport phenomena, see \reqs{final} and Fig.~\ref{fig1}, resembling those in superconductors \cite{LarkinOvchinnnikov75,LarkinOvchinnikovReview}.
 The confirmation of the peak effect in the
dissipation
and the threshold behaviour in the drift
provides evidence for the existence of the vortex core reconstruction and further our understanding of the dissipative vortex transport.

We thank William Glaberson and Nadav Katz for informative discussions.
This research has been supported by the United States-Israel Binational Science Foundation (BSF) grant No. 2012134 and the Israel Science Foundation (ISF) grant No. 302/14 (O.A.) and by Simons foundation (I.A.).


\begin{thebibliography}{99}
\bibitem{potential} It is natural to assume $r_c \gtrsim \xi$ as the weak disrodered potential is linearly screened
by the bosons which suppresses  $q\gtrsim 1/\xi$. The additional reason for this assumption is the averaging
of the density in the vortex core.
\bibitem{footnote1}
 The expressions for the energies $\epsilon_{d,q}$ in terms of the parameters of the model, see \reqs{eq:1} -- \rref{eq:4} are
\[
\begin{pmatrix}
\epsilon_d
\\[4pt]
\epsilon_q
\end{pmatrix}
=\frac{\hbar^2}{2mr_c^2}
\begin{pmatrix}
\gamma^{2/3}\left(\frac{M_v}{m}\right)^{1/3}\left[\frac{\pi}{4}\int dx\,x^2 G(x)\right]^{2/3}\\[4pt]
\gamma^{2}\left(\frac{M_v}{m}\right)^{3}\left[\frac{\pi}{2}\int dx G(x)\right]^{2}
\end{pmatrix}
\]
the condition for the $\chi_E$ peak to be within applicability of our theory, $\epsilon_d < \frac{\hbar^2}{m\xi^2}< \epsilon_q$,
restricts the disorder srength $(n_0\xi^2)^{-3/2}<\gamma<(n_0\xi^2)^{-1/2}$ which seems realistic.
\bibitem{Sorry}We will not write terms for the elastic scatterings of phonons on the vortex which
do not change our results qualitatively. We also omitt logarithmic energy dependence in the votex density of states.
It implies that with logarithmic accuracy, $2\epsilon(p)\approx p\partial_p\epsilon(p)$, and all the logarithms in the integrals are
treated as constants with the energy argument of the logarithm replaced with the relevant energy scale.
\end{thebibliography}
\end{document}